# NUMERICAL MODELS AND OUR UNDERSTANDING
# OF ASPHERICAL PLANETARY NEBULAE *

Garrelt Mellema[1] and Adam Frank[2]

[1] Dept. of Mathematics, UMIST, P.O. Box 88, Manchester M60 1QD, UK

[2] Dept. of Astronomy, Univ. of Minnesota, 116 Church St., Minneapolis, MN 55455, USA

astro-ph/9410057   19 Oct 1994

## ABSTRACT

The status of numerical hydrodynamical models for Planetary Nebulae is re-
viewed. Since all of the numerical work is based on the interacting winds model,
we start out with a description of this model and give an overview of the early
analytical and numerical models. Subsequently we address the numerical mod-
els which include radiation effects, first of all the ones which neglect any effects
of stellar evolution. These 'constant environment' models are shown to closely
match typical observed nebulae, both in images and kinematic data. This shows
that the basic generalized interacting winds model gives a good description of
the situation in aspherical PNe. Next we discuss models that do include the
effects of stellar and fast wind evolution. This introduces several new effects, the
most important of which are the formation of a surrounding attached envelope,
and the modification of the expansion of the nebula, which helps in creating
aspherical PNe very early on in their evolution. The ionization of the slow wind
also leads to a gradual smoothing out of its aspherical character, working against
aspherical PNe forming in later stages. Finally we discuss some applications of
the model. These are the predicted X-ray map, and possible explanations for
temperature fluctuations and hot haloes.

## 1.   INTRODUCTION

The now generally accepted model to explain the formation of Planetary Nebulae (PNe)
is what is known as the Interacting Stellar Winds model or ISW model. An important factor in
the success of this model is the collection of results from numerical modelling of PN formation
based on ISW. These results show that the ISW model can explain many if not all the properties
of these objects. This review deals with these numerical models.

---

The last few years saw a lot of activity in the field of numerical radiation-gasdynamics of PN formation, as is illustrated by the fact that no less than three theses on this subject were published[1,2,3]. Beside PNe, in the ISW model has recently been applied to other astrophysical systems, such as surroundings of the precursor of SN1987a[4], symbiotic stars[5], Wolf-Rayet stars[6], classical novae[7], eta Carinae[8], etc. In this review we will limit ourselves to the PN case.

## 2.   BACKGROUND

The ISW model was first put forward as a possible way of explaining PN formation by Kwok and collaborators in 1978[9]. Previously it had been applied to bubbles around massive stars[10-12]. Mainly to define the jargon, we now briefly describe the basic ISW model. The PN is supposed to form from the interaction between a slow wind (i.e. material lost during the AGB-phase in a Mira wind and/or a superwind) and a (post-AGB) fast wind. This fast wind sweeps up a shell of material in the slow wind. This swept-up shell is identified with the actual bright nebula and is bounded by an outer shock on the outside and a contact discontinuity on the inside. Beyond this contact discontinuity is a large volume of hot, tenuous, shocked fast wind material, known as the hot bubble. This region is bounded on the inside by the so called 'inner shock', beyond which is the actual fast wind.

Going back to work on aspherical nebulae around massive stars[13], it was quickly realized that if the slow wind has an axi-symmetric density distribution, this might well explain the many aspherical PN morphologies[14-18]. This extension of the original (spherical) ISW model we refer to as the Generalized ISW (GISW) model.

Numerical modelling of PN formation goes back to the work of Mathews in the sixties who studied the then generally accepted 'sudden ejection' model for PNe[19]. The ISW model suggestion triggered a new series of numerical studies[20-22]. All of these considered the case of perfectly spherical nebulae and included varying amounts of radiation physics.

The publication of the analytical studies of the GISW model mentioned above, was soon followed by the first numerical studies[23-26]. All of these models assumed perfect cylindrical symmetry and no energy losses or gains due to radiation processes. The numerical results nicely confirmed the analytical predictions and indicated that the GISW model might indeed explain a majority of observed PN shapes.

However in real PNe heating and cooling due to radiative processes play an important role in the energy budget and a proper comparison between observations and theory requires

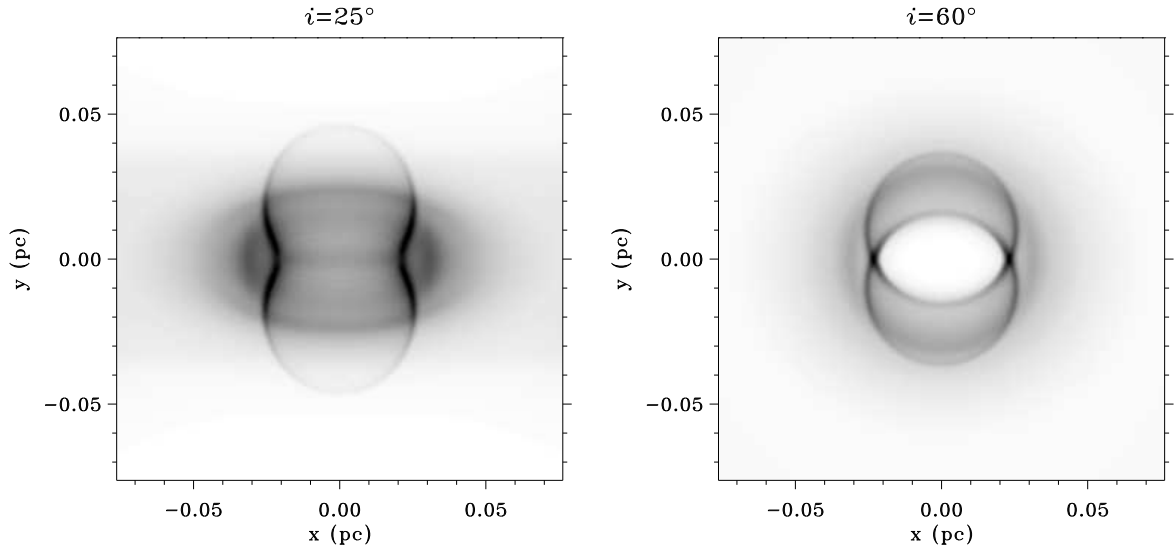

Figure 1. *Two Hα images of the same model nebula (of bipolar type morphology), seen at incli-
nations 25° and 60°. Notice the difference in apparent morphology.*

the simultaneous treatment of the gasdynamics and the radiation physics. For the heating and
cooling rates to be calculated properly one needs to know the ionic abundances of the relevant
species and these in turn depend on the ionization and recombination rates. What one basicaly
needs is a simple photo-ionization calculation in step with the gasdynamics.

Frank & Mellema developed such a method[27]. In it heating of the gas takes place through
photo-ionization of H, He, and He$^+$, and cooling through recombination of H and He, and col-
lisional excitation of H, He, He$^+$, C$^{3+}$, N$^+$, O$^+$, and O$^{2+}$. To calculate the necessary ionic
abundances, ionizations through photons and collisions and radiative recombinations are consid-
ered. To follow the transfer of radiation they use the 'on the spot' approximation. The stellar
spectrum is assumed to be blackbody. A similar, but more detailed method is used by Marten in
his spherical models[3,28−31].

Using this method, Mellema & Frank published models for the formation of aspherical
PNe[32−34]. In these models the effects of stellar evolution are neglected, that is the properties
of both the central star and its fast wind are assumed to be constant. The calculation of the
cooling allows the construction of synthesized observations (both images and long slit spectra)
from the results of the simulations. The synthesized images and spectra show very good agreement
with typically observed nebulae. The synthesized images also show how different morphological
features can be explained by one model nebula, seen at different inclinations (see Fig. 1 for an
example).

It is important to note that only with these radiative models realistic images and spectra can be constructed, and the claims of success for the GISW model from the non-radiative simulations be substantiated. In particular, a close comparison between radiative and non-radiative simulations shows significant differences in the density and temperature structure of the model nebulae[34].

Although the central star does not evolve in these models, an evolutionary effect did show up. After the slow wind has been ionized, its initial asphericity starts to be smoothed out on a time scale determined by the local sound speed ($\sim 10 \; \mathrm{km\,s^{-1}}$). For a distance of about $10^{15}$ m from the central star, this time scale is about 3000 years. The result of this effect is that slowly expanding nebulae which take substantially longer than this time to reach a size of $10^{15}$ m, are expected to be fairly spherical. This may be the reason why many older nebula are observed to be roughly spherical.

## 3.   EVOLUTION

A typical post-AGB star evolves considerably on a time scale of about 1000 years, and the fast wind is expected to change with the central star. Therefore realistic models should take into account the effects of changes in the stellar and fast wind properties.

Models for spherical nebulae which consider stellar evolution go back to Schmidt-Voigt & Köppen[22], and more recent models were published by Marten[28−31], Frank[35], and Mellema[36]. For aspherical nebulae results were published by Mellema[2,38].

The addition of these evolutionary results greatly increases the complexity of models, making even the relatively simple spherical models yield interesting new results. Many examples of this can be found in the papers quoted above. In this section we will concentrate on two effects which are particularly relevant for the nebular morphology. Both are caused by the gradual hardening of the stellar spectrum during the central star's traverse across the HR diagram (see Schönberner's contribution in these proceedings). The two effects are

 ◇ Envelope formation

 ◇ Effects on the nebular expansion

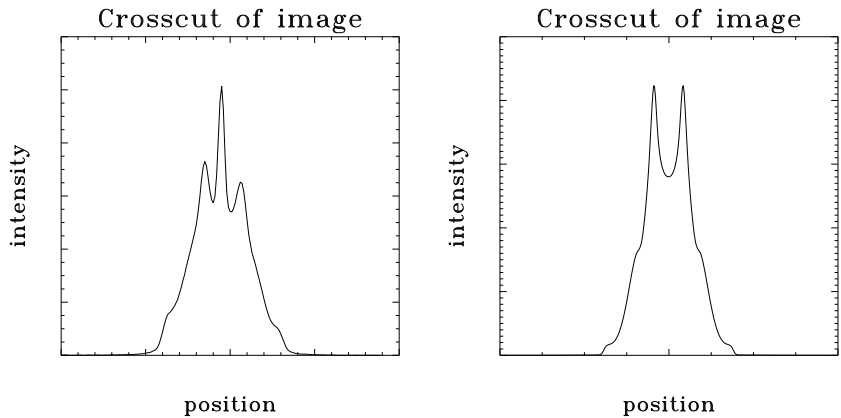

*Figure 2. Cross cuts across [OIII] images of NGC 6826 (left) and a model nebula (right). Notice the similarity in the profile of the surrounding envelope. The central peak in the NGC 6826 image is its central star.*

Envelope formation was first reported by Marten et al.[29] and was studied in more detail by Mellema[36,37] It is caused by the dynamical influence of the ionization D-front which forms in the early stages of nebular formation. This D-front slowly moves through the slow wind and in the process sweeps up a shell and when the stellar spectrum hardens, the ionization front breaks out and rapidly ionizes the rest of the slow wind. The shell that was formed by the front persists even after the front itself has disappeared. When one constructs the images from the models, this shell shows up as a diffuse attached envelope around the bright core nebula. The core nebula is of course formed in the usual way through the interaction between the fast wind and the slow wind.

The envelopes formed in the models closely match the observed ones (see the example in Fig. 2). These envelopes are sometimes called attached haloes or inner shells in the observational literature. Some examples are NGC 3242, IC 3568, NGC 6826. Typically observed properties such as the relative size and surface brightness, the linear emission profiles, and the expansion velocities are all reproduced by the models. In particular the initially surprising fact that in some objects the envelopes are seen to be expanding faster than their core nebulae[38], is naturally explained when the envelopes were formed by the ionization D-front[36].

Note that this formation mechanism for the attached envelopes implies that one should be very careful in deriving any characteristics of AGB mass loss (such as variations and their time scales) from these inner parts of the PN.

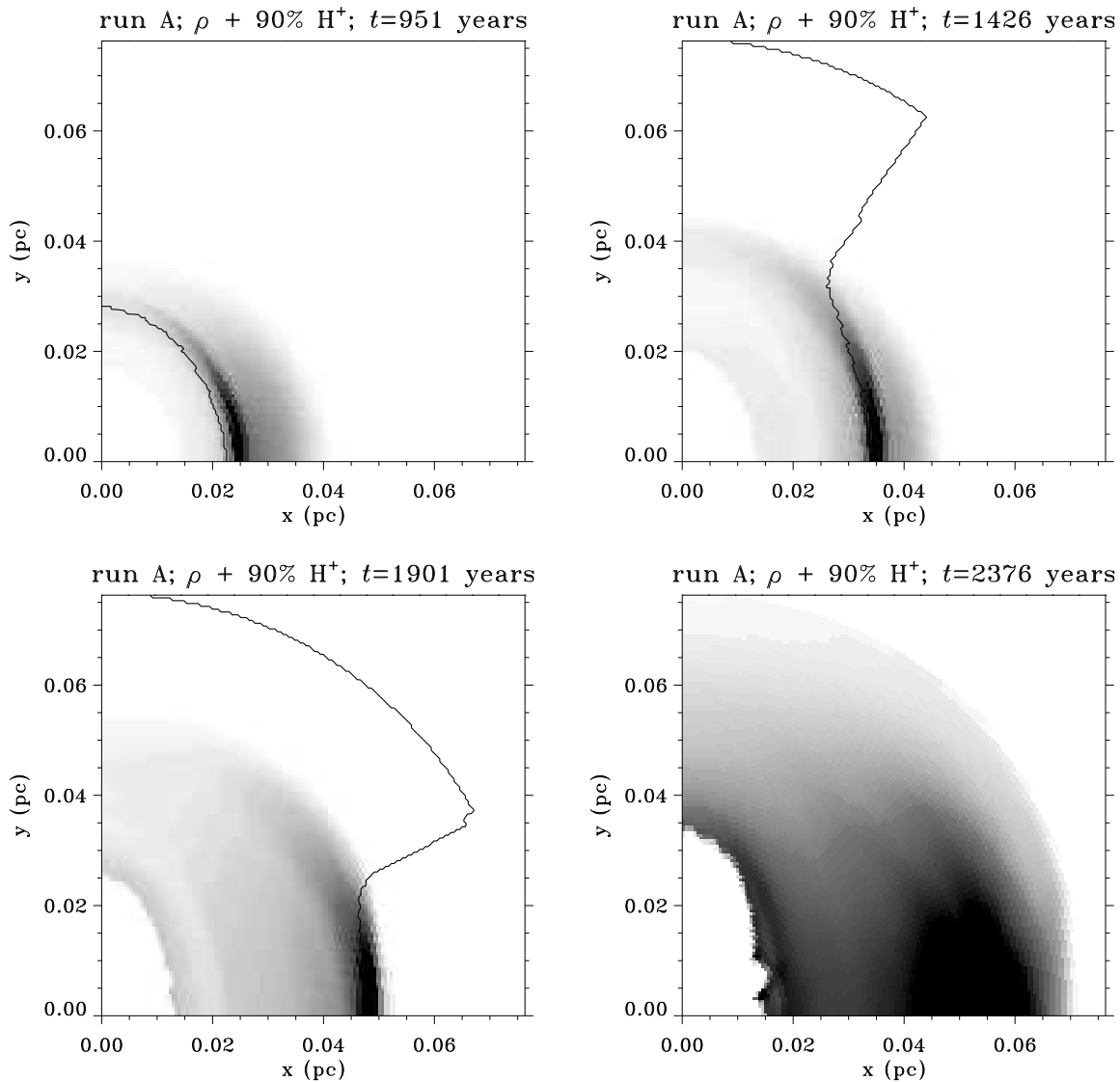

*Figure 3. Greyscale representation of the density of an evolving nebula, darker shades represent higher values. The dark line represents the ionization front. The boxes show the situation at four consecutive times. Due to the differential pressure increase caused by the ionization of the slow wind, the nebula expands in the polar but not in the equatorial direction.*

The second important effect introduced by the ionization of the slow wind is a modification of the expansion of the nebula. As the inner parts of the slow wind get ionized, the pressure there increases. This makes it harder for the fast wind to push up a shell and the effects is that upon ionization, the nebular expansion either slows down, stalls, or even reverses, depending on the ram pressure of the fast wind and the density of the slow wind. Only the slow wind density plays a role because the balance between heating and cooling will fix the slow wind temperature.

This also means that when the slow wind density varies, as it does in the GISW model,

the effect will be different at different positions in the nebula. The expansion will slow down much more in the denser equatorial parts than in the less dense polar part. The effect of this is that an aspherical nebula can form quite early on in the PN phase, right after ionization starts. Comparing the numbers for one case, Mellema[2] (page 153) finds that this 'ionization shaping' makes the asphericity increase twice as fast as in the equivalent constant environment situation, see Fig. 3

In the later stages, as the fast wind velocity increases, the nebular expansion velocity increases again. This means that in these later stages the nebulae are smaller than would follow from their expansion velocity at that time, or in other words, the derived expansion ages are less than the real ages.

## 4.    APPLICATIONS

At this point the numerical models seem to be producing fairly realistic nebulae. This opens up the possibility of using the models to study nebular problems. The standard equilibrium models for PNe have left a number of phenomena unexplained, and it may be that the dynamic models can offer an explanation. Here we want to point out three of these 'applications'

1. Soft X-ray emission
2. Temperature fluctuations
3. Hot haloes

### 4.1    Soft X-ray emission

In order to approximate the soft X-ray emission from the model nebula, we calculated the total bremsstrahlung flux between 0.5 and 2 keV. Figure 4 shows the H$\alpha$ and soft X-ray image of a particular model. The soft X-rays are mainly originating from a thin layer just inside the optical image. The total energy content in the soft X-rays is only a small fraction of the fast wind energy input ($< 0.01\%$). Only this interface efficiently produces soft X-ray emission is because the rest of the hot bubble is too hot.

The actual thickness of this layer depends on the amount of thermal conductivity between the nebula and the hot bubble. A predominantly toroidal magnetic field (which is what is expected[39,40]), will reduce the amount of conductivity drastically.

Observations of soft X-rays from PNe are not of high enough resolution to actually study the distribution of the radiation[41] (Kreysing et al. 1992), but in the related case of Wolf-Rayet nebulae observations do point to only a thin soft X-ray producing layer[42]. Our simple analysis does not produce an X-ray spectrum for comparison to the ROSAT results, see Zweigle's contribution in these proceedings for this.

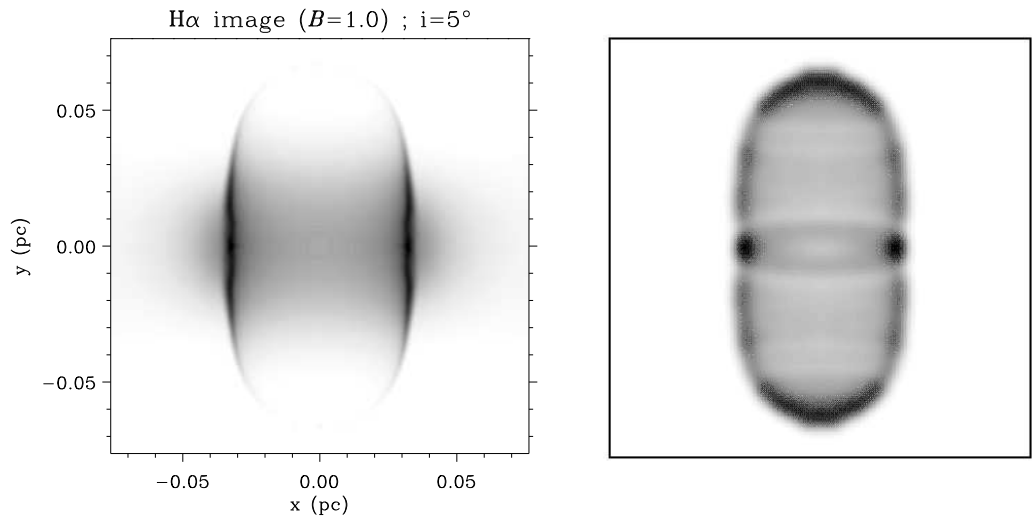

Figure 4. *The Hα image (left) and the soft X-ray image (right) of the same model nebula (of the elliptical type), seen at inclinations 5°. The soft X-ray image has been smoothed.*

*4.2    Temperature fluctuations*

In classical nebular analysis, different electron temperatures are found when using different methods (for instance from [OIII]5007 line ratios and the Balmer jump). This is known as the problem of 'temperature fluctuations'[43,44]. It is still unclear what causes this. Are there real temperature fluctuations and if yes, on which scale? Are the different methods measuring different (large) regions in the nebula, or is the material distributed in small clumps with varying densities and temperatures? The numerical models can be used to check some of these possibilities.

As a preliminary study one can look for real variations in the electron temperatures. It turns out that nebular material which is in thermal equilibrium does not show temperature variations. This should not come as a surprise, since static photo-ionization studies never showed any temperature fluctuations. However, some parts of the nebula can be out of thermal equilibrium because of the outer shock moving through. The cooling distance behind the outer shock depends on the square of the density. This means that in the case of aspherical nebulae the cooling region can be very small in the dense regions and quite extended in the less dense regions, resulting in some areas of the nebula being almost entirely in equilibrium and some out of it. When projecting the nebula on the sky these areas may overlap, suggesting temperature fluctuations. In some of our simulations with reasonable parameters for the slow wind density this behaviour is actually seen, showing that for the PN case this is in fact happening.

A better way of finding out whether these effects result in observed temperature fluctuations is to calculate the forbidden line ratios from the models and calculate the temperatures and densities from those. If these show similar fluctuations to the observed ones, the explanation becomes plausible. We plan to carry out a study like this in the very near future.

*4.3    Hot haloes*

In some nebulae, the surrounding haloes are observed to have high electron temperatures. Reported cases are NGC 6543, NGC 6826, and NGC 7662[45−48]. Static photo-ionization codes fail to produce an explanation for this and intricate shock heating mechanisms have been put forward to explain this phenomenon[47,49].

Numerical models that use an evolving central star turn out to produce hot haloes in a rather straightforward way. This was first reported by Marten[30], but we find exactly the same effect in our models. As is well known from photo-ionization studies, due to the $\nu^{-3}$ dependence of photo-ionization cross section, the radiation field is much harder near to the ionization front. In the relatively low density haloes this has the effect that as the ionization front moves through, the gas is heated to higher than equilibrium temperatures ($\sim 20\,000$ K) and takes a while to cool down. This time depends on the density. The case reported in Marten (1993) has a cooling time of about 200 years, whereas in our model, which has a lower slow wind density, this is 1200 years. During this cooling time the halo would be observed to have a high temperature. Whether or not this is the true explanation for the observed hot haloes depends on their density, which should support relatively long cooling times.

## 5.    CONCLUSIONS

• The GISW model is well established as the explanation for aspherical PNe. The increased complexity of the numerical models has only improved the correspondence between observations and the models.

• Models which include the stellar evolution introduce interesting new effects:
  ◇ Formation of surrounding envelopes, which match the observed ones quite closely.
  ◇ Modification of expansion by ionization of slow wind. For the aspherical case this increases the asphericity of the nebula in the early stages.

⋄ Ionization of the slow wind leads to a gradual erosion of its asphericity with a typical time scale of about 3000 – 5000 years.

● Since they provide the complete nebular structure, the models can be used to study typical nebular problems. We presented three examples:

⋄ The soft X-ray image. The emission is expected to come from a thin area just inside the optical nebula, and represents a very small fraction ($< 0.01\%$) of the fast wind energy content.

⋄ A search for the explanation of observed 'temperature fluctuations'. This might be caused by part of the nebula being out of thermal equilibrium.

⋄ An explanation for the observations of 'hot haloes'. These can be explained by high temperatures as a result of recent ionization.

## ACKNOWLEDGMENTS

We thank Bruce Balick and Vincent Icke as a continuing source of inspiration and advice, and Noam Soker for inviting us to write this review.

## FOOTNOTES AND REFERENCES